\documentstyle[12pt,psfig]{article}

\begin{document} 

\title{Final State Interactions in the
Near-Threshold Production of Kaons from Proton-Proton
Collisions\thanks{Supported 
by Forschungszentrum J\"ulich}}
\author{A. Sibirtsev\thanks{On leave from the Institute of 
Theoretical and Experimental Physics, 117259 Moscow, Russia.} \
 and  W. Cassing \\
Institut f\"ur Theoretische Physik, Universit\"at Giessen \\
D-35392 Giessen, Germany}
\date { }
\maketitle
\vspace{-9.5cm}
\hfill UGI-98-7
\vspace{9.5cm}

\begin{abstract}
We analyse the $pp \to p \Lambda K$ cross section
recently measured at COSY arguing that the enhancement of the
production cross section at energies close to the
reaction threshold should be due to the
$\Lambda p$ final state interaction. We find that the experimental
$\Lambda p$ elastic scattering data as well as the
predictions from the J\"ulich-Bonn model are in reasonable
agreement with the new results on $K^+$-meson production.
We propose to study directly the final state interaction
by  measurements of the cross section as a function of the hyperon
momentum in the $\Lambda p$ cm system. 
\end{abstract}

\vspace{1cm}


\newpage

Recently, the COSY-11~\cite{Balewski} and COSY-TOF 
Collaborations~\cite{Bilger} measured the 
$pp\to p \Lambda K^+$ cross section at energies close to 
the reaction threshold. The experimental $K^+$ yield found is
twice as large as predicted in Ref.~\cite{SiCa} and by
F\"aldt and Wilkin~\cite{Faldt1} for excess energies 
$\epsilon < 7$~MeV whereas both models reproduce the data for  
$\epsilon > 50$~MeV. Here we will argue that this enhancement is due to
$\Lambda p$ Final State Interactions (FSI).

In Ref.~\cite{SiCa} the total cross section for the 
$pp\to p \Lambda K^+$ reaction has been calculated 
within the One-Boson-Exchange (OBE) model including both the pion and 
kaon exchanges in line with the analysis performed at 
high energies~\cite{Kaon}. In the latter work 
the  FSI between the $\Lambda$-hyperon and the proton 
has been neglected for reasons of simplicity but also 
due to the large uncertainty  in the low energy 
$\Lambda p$ scattering cross section~\cite{Alexander,Lambdap}.
The free parameters of the model, i.e.  the coupling constants and the 
cut-off parameters of the form factors for the $NN\pi$ and $NYK$ 
vertices, were fitted to the experimental data 
taken at high energies~\cite{LB}. 

On the other hand,
F\"aldt and Wilkin~\cite{Faldt1} calculated the energy dependence of 
the $pp\to p \Lambda K^+$ cross section by using only the one pion 
exchange in line with the calculations from 
Ref.~\cite{Pion}.  Moreover, it was  assumed that the production 
amplitude is constant and is related to the $pp\to pp\eta$
reaction. The FSI between the $\Lambda$-hyperon and the  
proton then was incorporated via the effective range approximation.
With only a single free parameter fixed by the $\eta$ production
data, the calculations from Ref.~\cite{Faldt1} reasonably 
reproduced the $pp\to p \Lambda K^+$ cross section at
$\epsilon =2$~MeV~\cite{Balewski1} available at that time.
Note, however, that the FSI correction factor at this energy
is about $\simeq 14$ and therefore the production amplitude due
to the pion exchange itself is very small. On the other hand,
Tsushima et al.~\cite{Tsushima} 
calculated the  $pp\to p \Lambda K^+$ production 
amplitude in a microscopic model which illustrates that near 
threshold the pion exchange contribution to the reaction is small and 
underestimates the most recent data from COSY~\cite{Balewski,Bilger}   
by about a factor  $\simeq 10$. Our present work is to clarify these
partly conflicting results. 

We recall that the total cross section for the 
reaction  $pp\to p \Lambda K^+$ is obtained by 
integrating the differential
cross section $d^2 \sigma/dt ds_1$ over the available phase space, 
\begin{equation}
\label{xsection}
\sigma = \int dt ds_1 \frac{d^2\sigma}{dt ds_1} = 
\frac{1}{2^9 \pi^3 q^2 s} 
\ \int dt ds_1 \ \frac{q_K}{\sqrt{s_1}} \ |M(t, s_1)|^2 . 
\end{equation}
Here $s$ is the squared invariant mass of the
colliding protons, $q$ is the proton momentum in
the center-of-mass while $t$ is the squared 4-momentum 
transfered from the initial proton to the final hyperon
or proton in case of  the kaon or pion exchange,
respectively. Moreover, $s_1$ is the squared invariant mass
of the $Kp$ or $K \Lambda$ system, respectively, while $q_K$ is the
kaon momentum in the corresponding center-of-mass system.
In Eq.~(\ref{xsection}) $|M|$ is the amplitude of
the reaction which is an analytical function of $t$ and $s_1$.

\begin{figure}[h]
\psfig{file=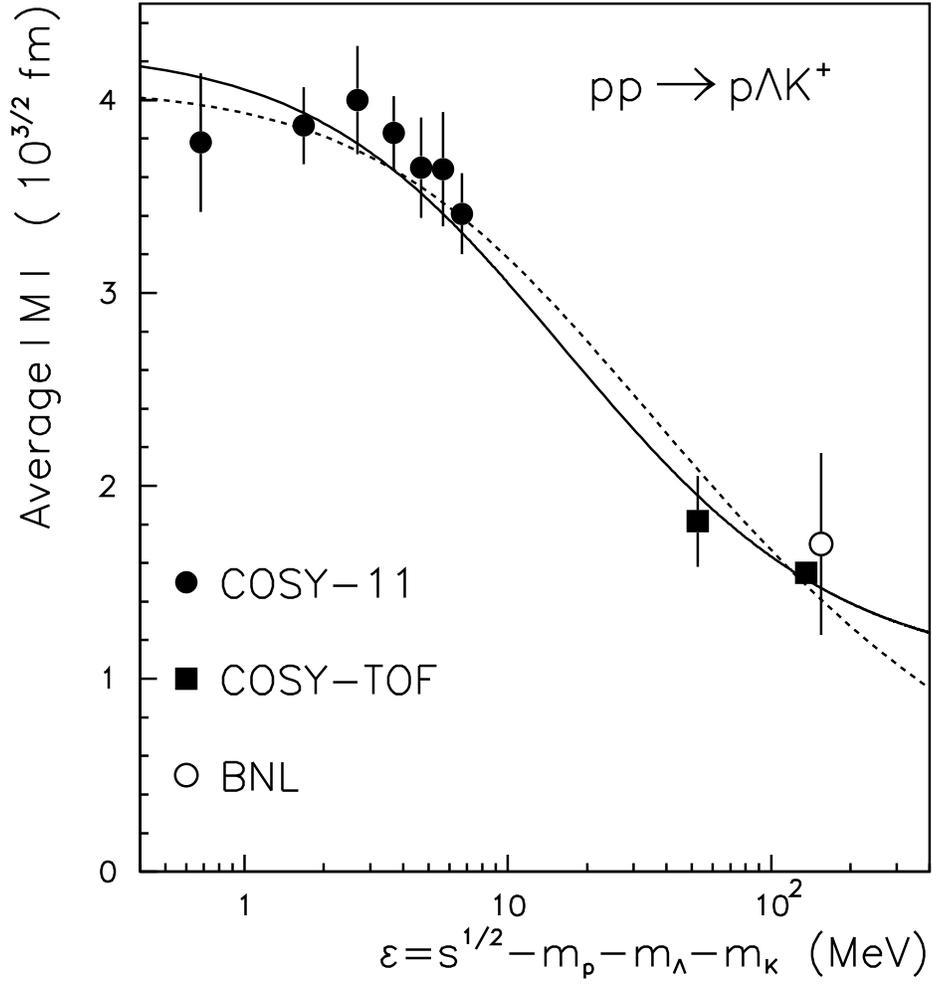,width=15cm}
\caption[]{\label{cosy1}The average amplitude for the
$pp\to p \Lambda K^+$ reaction
as a function of the excess energy. The symbols
show the results extracted from the experimental 
data of Refs.~\protect\cite{Balewski,Bilger,LB}. The lines are our 
calculations with FSI as described in the text.}
\end{figure}

Let us start with the experimental $pp\to p \Lambda K^+$ 
cross section  and extract the reaction amplitude averaged
over $t$ and $s_1$ by taking $|M|^2$ out of the integral in
Eq.~(\ref{xsection}). The average reaction amplitude then can be
determined by comparing $\sigma$ from Eq.~(\ref{xsection}) with the
corresponding experimental cross section from 
Refs.~\cite{Balewski,Bilger,LB}. The results for 
the average matrix element $|M|$
are shown in Fig.~\ref{cosy1} as function of the excess energy
$\epsilon = \sqrt{s}-m_p-m_{\Lambda}-m_K$, with $m_p$,
$m_{\Lambda}$ and $m_K$ being the mass of the proton, 
$\Lambda$-hyperon and kaon, respectively. Obviously, the matrix
element $|M|$ as evaluated from the data is not constant and decreases
substantially with the excess energy $\epsilon$.

Following the Watson-Migdal approximation the total reaction 
amplitude can be factorized in terms of the production
$|M_{prod}|$ and FSI amplitude $|A_{FSI}|$. Since at high
energies the FSI is  negligible, the latter amplitude 
should converge to 1 for $\epsilon \rightarrow \infty$. 
Within the OBE model the squared production amplitude, for instance
for the pion exchange, is given 
as~\cite{SiCa,Kaon,Pion}\footnote{One should add the amplitude
corresponding to the exchange graph.}
\begin{equation}
\label{prod}
|M_{prod} (t, s_1)|^2= g^2_{NN\pi} \ \frac{t}{(t-\mu^2)^2} \
{\left\lbrack \frac{ {\Lambda}_{\pi}^2-{\mu}^2 }
{{\Lambda}_{\pi}^2-t} \right\rbrack}^2 \ 
|A_{\pi^0 p\to \Lambda K^+}(s_1)|^2 ,
\end{equation}
where $g_{NN\pi}$ is the coupling constant and $\mu$ is the pion
mass. The form factor of the $NN\pi$ vertex is given in
brackets with ${\Lambda}_{\pi}$ denoting the cut-off parameter.
In Eq.~(\ref{prod}) $|A|$ is the amplitude for the
reaction $\pi^0 p\to \Lambda K^+$, which can be calculated
microscopically within the resonance model\footnote{Since
the resonance properties are fitted to the experimental data,
both the resonance model and Eq.~(\ref{piNKL}) should give 
the same results.} or evaluated from the experimental 
data as
\begin{equation}
\label{piNKL}
|A_{\pi^0 p\to \Lambda K^+}(s_1)|^2 = 
16  \pi \ s_1 \ \frac{q_{\pi}}{q_K} \ 
\sigma_{\pi^0 p\to \Lambda K^+}(s_1) ,
\end{equation}
where $q_{\pi}$ is the pion momentum in the $\Lambda K$
center-of-mass system and $\sigma_{\pi^0 p\to \Lambda K^+}$
is the physical cross section. 

\begin{figure}[h]
\psfig{file=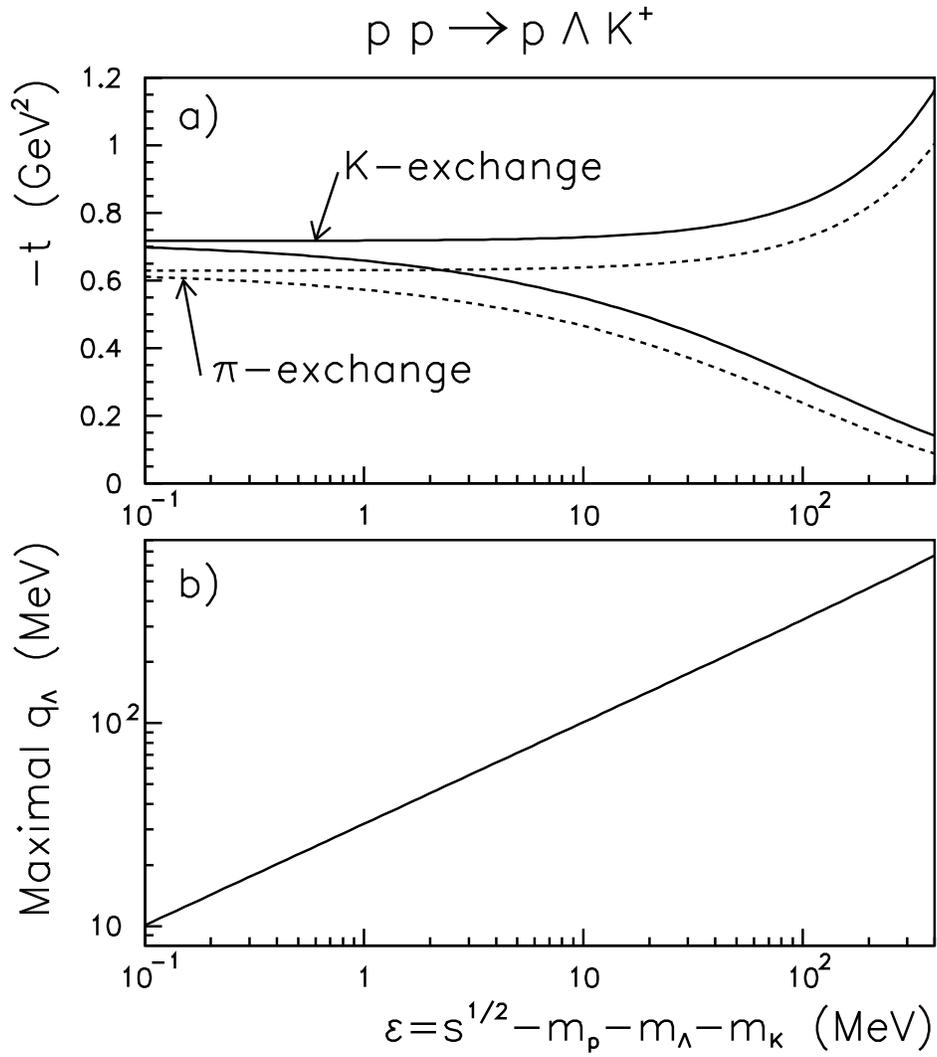,width=15cm}
\caption[]{\label{cosy5}The range of the 4-momentum transfer
squared (a) and maximal hyperon momentum in the $\Lambda p$
cm system (b) as function of the excess energy $\epsilon $.}
\end{figure}

Since  $m_{\Lambda}+m_K \le \sqrt{s_1} \le m_{\Lambda}+m_K +\epsilon $,
thus close to the $pp\to p \Lambda K^+$ reaction threshold
the amplitude $|A_{\pi^0 p\to \Lambda K^+}|$ is almost constant.
Moreover, the 4-momentum transfer squared $t$ is a slowly varying
function of energy at low $\epsilon $ as  shown
in Fig.~\ref{cosy5}a). Therefore it is a quite reasonable
approximation to assume that the production amplitude $|M_{prod}|$
is almost constant near the threshold. Similar arguments
can be set up for the kaon exchange amplitude.
We thus conclude that the deviation of the reaction amplitude
$|M|$ (shown in Fig.~\ref{cosy1}) from a constant
value is due to the FSI. 

\begin{figure}[h]
\psfig{file=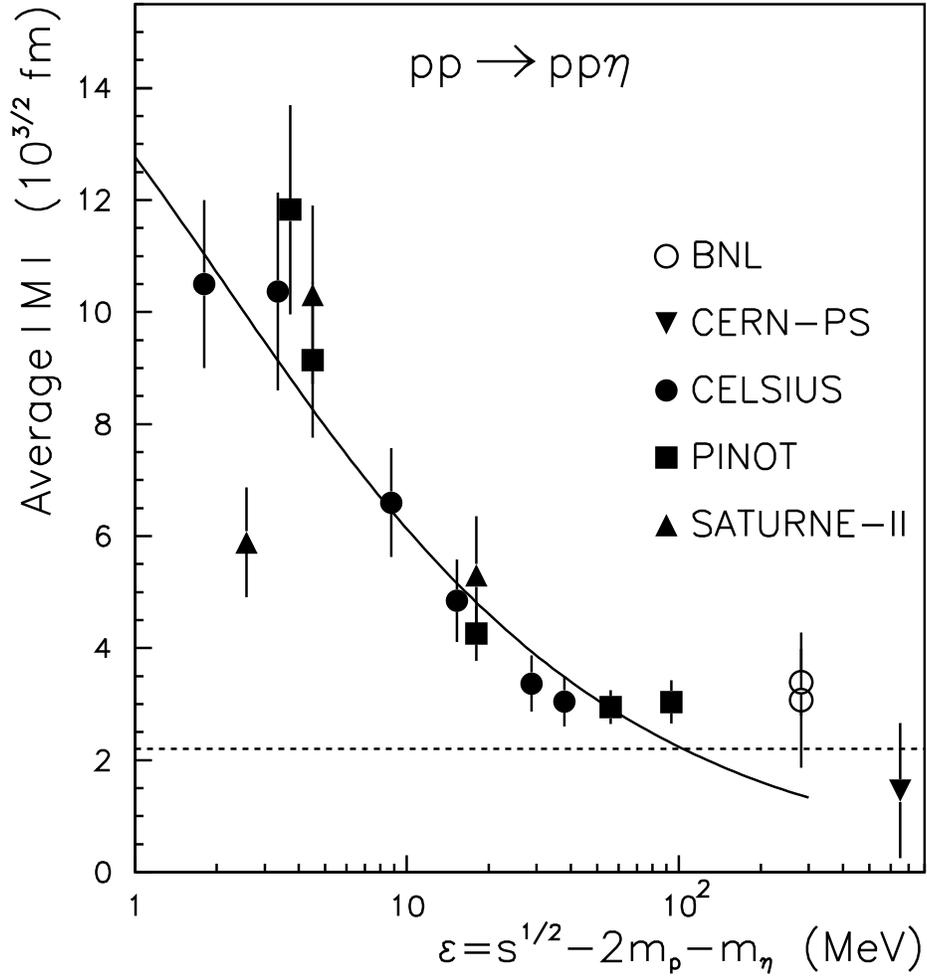,width=15cm}
\caption[]{\label{cosy2}The average amplitude for the
$pp\to p p \eta$ reaction. The symbols
show the results extracted from the experimental 
data~\protect\cite{Bergdolt}. The
dashed line shows the production amplitude while
the solid line is the total amplitude corrected by 
FSI~(\protect\ref{prescr}) with parameters in line with 
the $pp$ interaction.}
\end{figure}

A similar conclusion is obtained by 
analysing the $pp \to pp \eta$ reaction cross section~\cite{Bergdolt}
in the same way as illustrated in
Fig.~\ref{cosy2}. Moreover, the experimental data  indicate that 
for kaon production the FSI correction is substantially smaller than
for $\eta$-meson production. We note that in principle
one should account for the FSI between all
final particles produced in the reaction
$pp\to p \Lambda K^+$; here we assume that the
$\Lambda p$ interaction is much stronger than the $K^+p$
and $K^+\Lambda $ interaction, respectively. 

Following the original idea of Chew and Low~\cite{Chew}
the amplitude $|A_{FSI}|$ is related to the  on
mass-shell elastic scattering amplitude $|A_{el}|$
for the reaction $\Lambda p \to \Lambda p$, which can be 
calculated from the corresponding physical cross section in analogy 
to  Eq.~(\ref{piNKL}). The strength of the FSI depends upon 
the $\Lambda $-hyperon momentum in the $\Lambda p$ 
center-of-mass system and for fixed excess energy the
relative momentum $q_{\Lambda}$
extends  from zero to its maximal value as  shown in
Fig.~\ref{cosy5}b). Thus to calculate the total cross section
for the reaction $pp \to p \Lambda K^+$ at 
$\epsilon < 100$~MeV~\cite{Balewski,Bilger} one needs
to know the $\Lambda p$ scattering amplitude for  
$0 \ge q_{\Lambda} \le 400$~MeV/c.

\begin{figure}[h]
\psfig{file=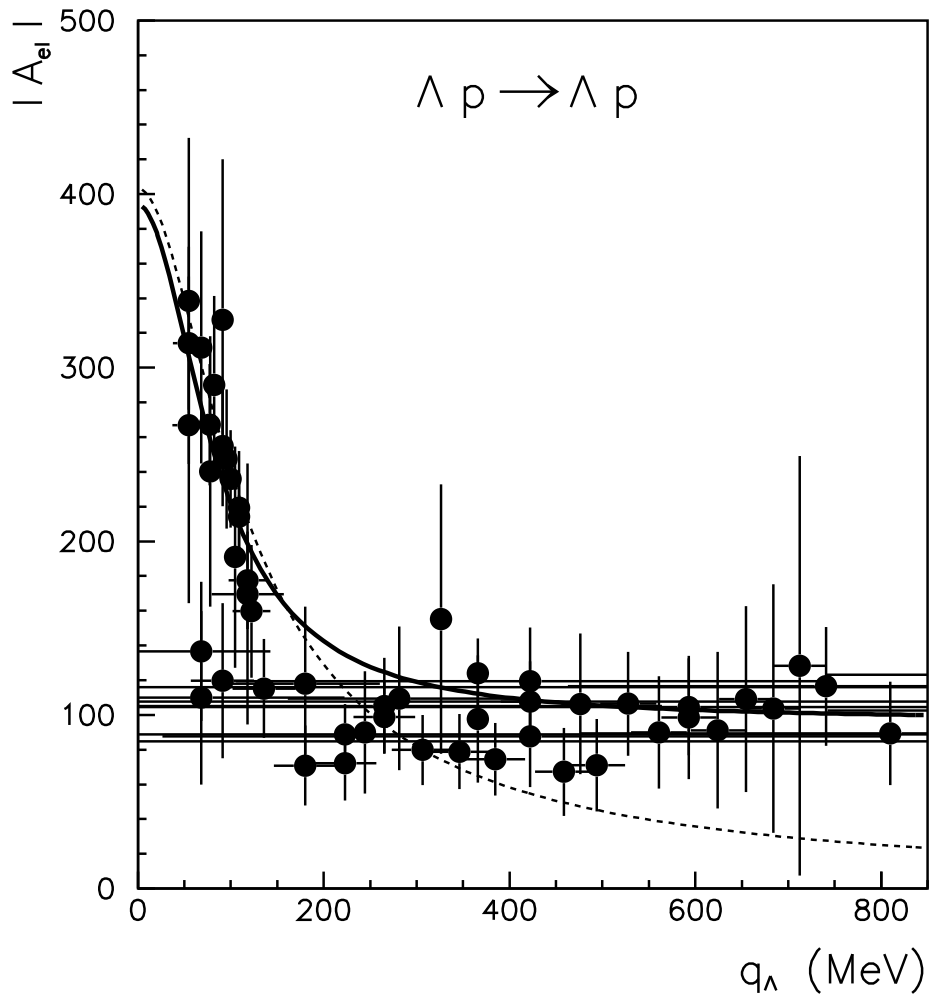,width=15cm}
\caption[]{\label{cosy4}The $\Lambda p$ elastic
scattering amplitude as a function of the hyperon momentum
$q_{\Lambda}$ in the center-of-mass system. The dots show the 
experimental data~\protect\cite{LB}; the fat solid line is our fit
while the dashed line shows the effective range approximation
with parameters from the J\"ulich-Bonn 
model~\protect\cite{Lambdap}.}
\end{figure}

Fig.~\ref{cosy4} shows the amplitude $|A_{el}|$ extracted from the
experimental data~\cite{LB}  plotted
as a function of $q_{\Lambda}$. Actually, there are no
data below 50~MeV/c and the experimental results have large 
error bars.
The amplitude $|A_{el}|$, furthermore,  can be calculated in the
hyperon-nucleon interaction model developed by Holzenkamp, 
Reuber, Holinde and Speth~\cite{Lambdap}.
Within the Bonn-J\"ulich approach the $\Lambda p$ elastic
scattering cross section is given in the effective-range 
formalism as
\begin{equation}
\label{efrange}
\sigma_{\Lambda p \to \Lambda p} =
\frac{\pi}{q_{\Lambda}^2+(-1/a_s+0.5r_sq_{\Lambda}^2)^2}+
\frac{3\pi}{q_{\Lambda}^2+(-1/a_t+0.5r_tq_{\Lambda}^2)^2},
\end{equation}
where $a$ is the scattering length and $r$ is the effective range,
while the indices $s$ and $t$ stand for singlet and triplet
$\Lambda p$ states. The scattering amplitude $|A_{el}|$
was calculated with $a$ and $r$ parameters
from Ref.~\cite{Lambdap} (model $\tilde {A}$) and  is
shown in Fig.~\ref{cosy4} by the dashed line.
Actually the effective range approximation is valid
at low energies and can not be extended to 
$q_{\Lambda} >200$~MeV/c. Following both the low energy
prediction from the Bonn-J\"ulich model and the experimental data
we can fit the $\Lambda p$ scattering amplitude as
\begin{equation}
\label{scat}
|A_{el}| = 
C \left\lbrack 
1+ \frac{\alpha \beta}{q_{\Lambda}^2+{\alpha}^2/4} \right\rbrack 
\end{equation}
with $\alpha = 170$~MeV, $\beta =130$~MeV and $C=87$. 
The result is shown by the fat solid line in Fig.~\ref{cosy4}.

The  FSI amplitude  now  is proportional to  the $\Lambda p$ 
elastic scattering amplitude, but normalized such that 
$|A_{FSI}| \to 1$ at large excess energies $\epsilon $. 
The solid line in Fig.~\ref{cosy1} shows our result
calculated with the $|A_{FSI}|$ averaged over the 
phase space distribution for $q_{\Lambda}$ and the production 
amplitude $|M_{prod}|= 1.05$~fm. The dashed line in 
Fig.~\ref{cosy1} illustrates the result obtained with the
prescription for FSI from Ref.~\cite{Faldt1,Faldt2} as
\begin{equation}
\label{prescr}
|A_{FSI}|^2 = \frac{2\beta_s^2}
{(\alpha_s+\sqrt{\alpha_s^2+2 \mu \epsilon } )^2}+
\frac{4\beta_t^2}
{(\alpha_t+\sqrt{\alpha_t^2+2 \mu \epsilon } )^2},
\end{equation}
where $\mu$ is the reduced mass in the $\Lambda p$ system
and the parameters $\alpha $ and $\beta $ were evaluated from the 
scattering length and the effective range for the singlet
and spin-triplet $\Lambda p$ interaction~\cite{Lambdap}.
Fig.~\ref{cosy1} illustrates a  good agreement between the
experimental data from COSY~\cite{Balewski,Bilger}
and our FSI approach and  proves the strong influence
of the $\Lambda p$ interaction in the final state.

\begin{figure}[h]
\psfig{file=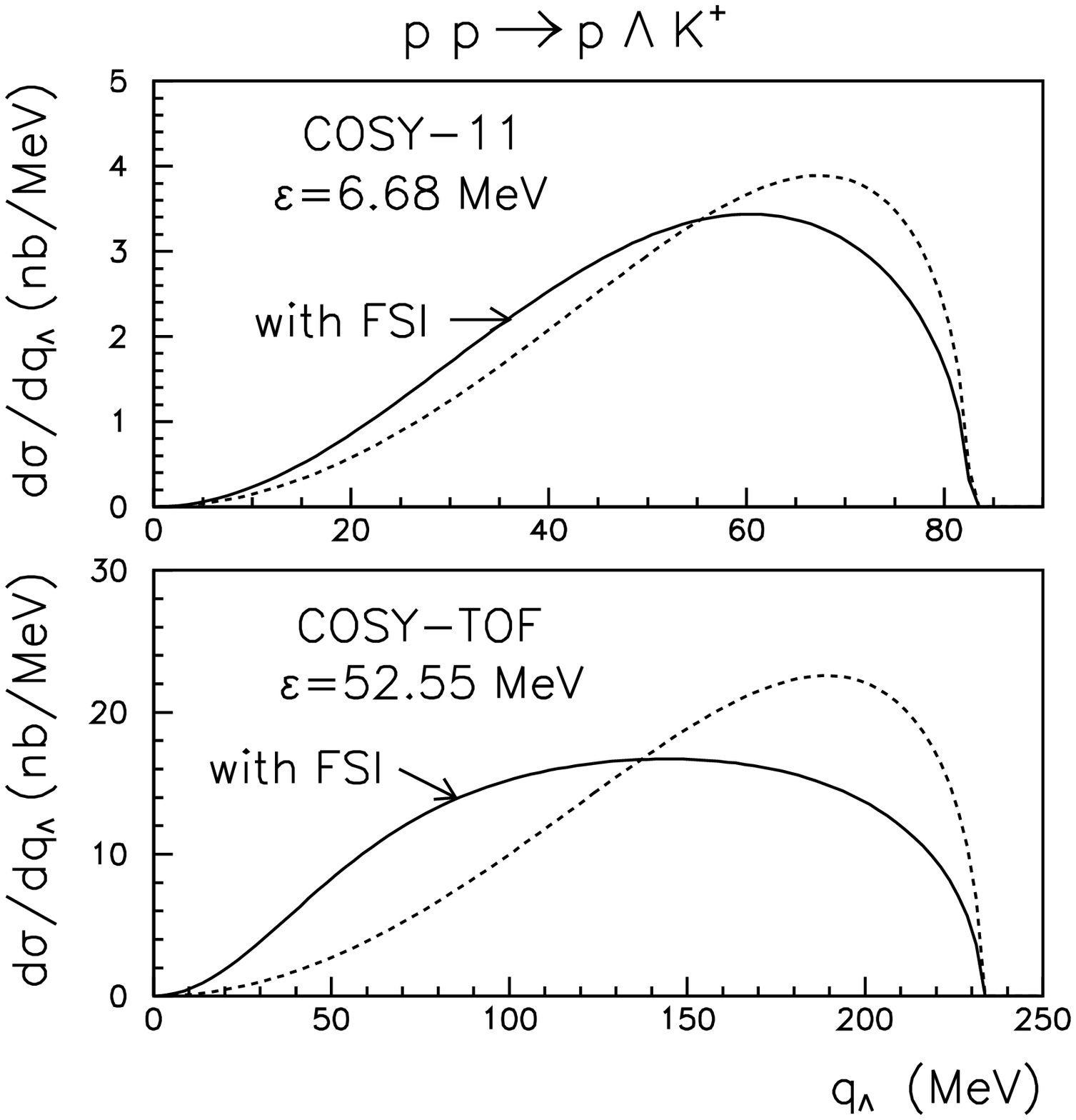,width=15cm}
\caption[]{\label{cosy6}The $pp \to p \Lambda K^+$
cross section as a function of the hyperon momentum in 
the $\Lambda p$ cm system calculated for two values
of the excess energy $\epsilon $. The solid lines show our results 
with  FSI correction while the dashed lines are the 
pure phase space distributions.}
\end{figure}

Actually, the FSI effect can be observed directly
by measuring the $pp \to p \Lambda K^+$ cross section as a
function of the momentum $q_{\Lambda}$. Fig.~\ref{cosy6}
shows this differential cross section 
calculated\footnote{The results are
normalized to the  experimental total cross section.}
by the phase space alone (dashed lines) and with 
FSI correction (solid lines) for the excess energies 
$\epsilon $ relevant for COSY-11 and COSY-TOF experiments.  
The enhancement at low  $q_{\Lambda}$ is due to the FSI and 
is more pronounced for the range of TOF energies.

We conclude, that the $pp \to p \Lambda K^+ $ cross section at 
energies close to the reaction threshold is strongly effected by
the hyperon-nucleon final state interaction. 
However, this effect is less pronounced as in the 
$pp\to pp \eta $ reaction close to threshold since the
$\Lambda N$ interaction is weaker than the $pp$ 
interaction at low relative momenta. Further experimental
studies are necessary for the direct measurement of the FSI
as a function of the relative momentum $q_{\Lambda}$ as well
as more upgrade calculations on the $Y p$ interaction,
which are under study in J\"ulich~\cite{Haidenbauer}.

\begin{thebibliography}{99}
\bibitem{Balewski}
        J.T. Balewski et al., preprint FZJ-IKP(I)-1997-1, 
	to be pub. in Phys. Lett. B.
\bibitem{Bilger}
        B. Bilger  et al., to be pub. in Phys. Lett. B. 
\bibitem{SiCa}
        A. Sibirtsev, Phys. Lett. B 359 (1995)  29;
        A. Sibirtsev and W. Cassing, Report No.~1787/PH of
	the H. Niewodnicza\'{n}ski Institute of Nuclear Physics,
	Cracow, Poland.
\bibitem{Faldt1}
        G. F\"aldt and C. Wilkin, Z. Phys. A 357 (1997) 241.
\bibitem{Kaon}
	E. Ferrari,  Nuovo Cim.  15 (1960)  652;
        J.M. Laget,  Phys. Lett.  B 259 (1991) 24;
        G.Q. Li and C.M. Ko, Nucl. Phys. A594 (1995) 439.
\bibitem{Alexander}
        G. Alexander et al., Phys. Rev. 173 (1968) 1452.
\bibitem{Lambdap}
        B. Holzenkamp, K. Holinde and J. Speth, 
	Nucl. Phys. A 500 (1989) 485;
        A. Reuber, K. Holinde and  J. Speth,
	Nucl. Phys. A 570 (1994) 543.
\bibitem{Pion}
        T. Yao,  Phys. Rev.  125 (1962)  1048. 
        J.Q. Wu and C.M.  Ko,  Nucl. Phys.  A 499 (1989) 810. 
\bibitem{LB}
        Landolt-B{\"{o}}rnstein, New Series, 
        ed. H. Schopper, I/12 (1988). 
\bibitem{Balewski1}
        J.T. Balewski et al., Phys. Lett. B 338 (1996) 859.
\bibitem{Tsushima}
        K. Tsushima, A. Sibirtsev and A.W. Thomas,
        Phys. Lett. B 390 (1997) 29; 
	nucl-th/9711028, to be pub. in Phys. Lett. B.
\bibitem{Bergdolt}
	A.M. Bergdolt et al., Phys. Rev. D48 (1993) R2969;
        E. Chiavassa et al., Phys. Lett. B 322 (1994) 270;
        H. Cal\"en et al., Phys. Lett. B 366 (1995) 39.
\bibitem{Chew}
        G.F. Chew and F.E. Low, Phys. Rev. 113 (1959) 1640).
\bibitem{Faldt2}
	G. F\"aldt and C. Wilkin, Phys. Let. B 382 (1996) 209.
\bibitem{Haidenbauer}
	J. Haidenbauer, private communication.
\end{thebibliography}
\end{document}